\pdfoutput=1
\documentclass[twocolumn,11pt]{article}
\usepackage[singlespacing]{setspace} 
\usepackage{graphicx}
\usepackage{epstopdf}
\usepackage{mathptmx}
\usepackage{microtype}
\usepackage{helvet}
\usepackage{courier}
\usepackage{placeins}
\usepackage{lineno} 

\usepackage[left = 2cm, top = 1.8cm, right = 2cm, bottom = 2cm]{geometry}

\usepackage[font=small,labelfont=bf]{caption}

\usepackage[nolist,nohyperlinks]{acronym}
\usepackage[colorlinks=true, linkcolor=black, citecolor=black, urlcolor=black]{hyperref}

\newcommand{\widthOnecol}{82mm}

\begin{document}

\begin{acronym}
	\acro{HRTF}{Head-Related Transfer Function}
	\acro{HRIR}{Head-Related Impulse Response}
	\acro{ILD}{Interaural Level Difference}
	\acro{ITD}{Interaural Time Difference}
	\acro{ANC}{Active Noise Control}
	\acro{RMS}{Root-Mean Square}
	\acro{SNR}{Signal-to-Noise Ratio}
	\acro{RECTF}{Residual Ear Canal Transfer Function}
\end{acronym}

\title{The Hearpiece database of individual transfer functions of an openly available in-the-ear earpiece for hearing device research}

\author{Florian Denk and Birger Kollmeier}
\date{\textit{Medizinische Physik \& Cluster of Excellence Hearing4All, Universit\"at Oldenburg, Germany}}
\maketitle

\subsection*{Abstract}
{
\textit{
We present a database of acoustic transfer functions of the Hearpiece, an openly available multi-microphone multi-driver in-the-ear earpiece for hearing device research. The database includes HRTFs for 87 incidence directions as well as responses of the drivers, all measured at the four microphones of the Hearpiece as well as the eardrum in the occluded and open ear. The transfer functions were measured in both ears of 25 human subjects and a KEMAR with anthropometric pinnae for five reinsertions of the device. We describe the measurements of the database and analyse derived acoustic parameters of the device. All regarded transfer functions are subject to differences between subjects as well as variations due to reinsertion into the same ear. Also, the results show that KEMAR measurements represent a median human ear well for all assessed transfer functions. The database is a rich basis for development, evaluation and robustness analysis of multiple hearing device algorithms and applications. The database is openly available at \url{https://doi.org/10.5281/zenodo.3733191}.
}
}


\section{Introduction}
Development and evaluation of hearing devices like hearing aids or hearables and appropriate algorithms is greatly facilitated by utilizing simulations. It is well understood that realistic simulations are required to obtain meaningful results \cite{grimm_toolbox_2019, pausch_extended_2018}. To simulate input signals of hearing devices, signals can be convolved with appropriate \acp{HRTF} that describe the acoustic free-field transmission to the hearing device microphone from a certain incidence direction. Several researchers have presented measurements of hearing device \acp{HRTF} \cite{pausch_extended_2018, denk_adapting_2018, oreinos_measurement_2013, kayser_database_2009}, and it has been shown that there are significant differences of the \acp{HRTF} between hearing device styles and microphone positions, as well as differences and perceptual consequences with respect to \acp{HRTF} measured in the unobstructed ear \cite{denk_spectral_2018, durin_acoustic_2014, hoffmann_quantitative_2013}. Also, the differences between individuals and the implications for designing signal processing algorithms has been recognized \cite{denk_adapting_2018, moore_personalized_2019}.
One limitation of existing datasets \ac{HRTF} is that while they can be utilized well to study the theoretical performance of algorithms, the authors usually used custom devices that are not available to other researchers. This means that others would have to build their own devices given the (often sparse) documentation in order to transfer their developments to real-time devices that are usable in the field.

Several other transfer functions related to hearing devices are crucial for their real-ear performance. However, they seem to be given little attention in current research, and the authors are not aware of public datasets of hearing device \acp{HRTF} that also include the driver responses, feedback paths, or transfer function to the occluded eardrum \cite{dillon_hearing_2012}. For instance, the responses of the driver(s) at the eardrum determine the sound that is perceived by the user, and recently several researchers tackled the problem of individualized equalization of the presented sound \cite{hoffmann_quantitative_2013, denk_equalization_2018, valimaki_assisted_2015}. Also, the feedback paths, i.e., the response of the driver at the device's microphones, is a factor that may greatly affect the performance of hearing devices especially if gain is to be provided \cite{spriet_feedback_2008}. Furthermore, many modern devices feature a non-occluding fit, i.e., significant sound energy from external sound sources enters the ear canal directly without being processed by the device, which has to be considered. Usually, the tightness of fit also interacts with the driver responses and feedback paths \cite{dillon_hearing_2012,blau_acoustics_2008}.

We here present a database of all linear transfer functions of the Hearpiece, a recently presented openly available in-the-ear earpiece with wired transducers for hearing device research \cite{denk_one-size-ts-all_2019}. The database contains on the one hand the \acp{HRTF} from 87 directions to the four microphones of the Hearpiece as well as the eardrum, both in the open ear and the ear occluded by the passive device. On the other hand, the responses of the two drivers at the eardrum as well as at the microphones integrated in the device (i.e., feedback paths) were measured. The transfer functions were measured in 25 human subjects and a KEMAR with anthropometric pinnae (G.R.A.S. 45BB-12 \cite{wille_iec_2016}) for each 5 reinsertions and variations of the sound field by placing a telephone near the ear. Furthermore, the between-device variation was assessed for 10 pairs of the Hearpiece that were either vented or completely occlude the ear. The database thus amounts to 169,878 \acp{HRTF} and 5740 driver responses.

This work includes a description of the conducted measurements and the database, as well as an evaluation of derived acoustic parameters of the Hearpiece. Note that while the general properties of the device were already analyzed in \cite{denk_one-size-ts-all_2019}, this work mainly focusses on the differences between subjects and devices from a series, variations of the transfer functions with reinsertion, and to what degree measurements in KEMAR are suitable to capture the acoustic properties of such an in-ear hearing device in a median human ear.
One special feature of the Hearpiece is a microphone in the ear canal, which allows the implementation of novel algorithms in the field of individualized equalization, feedback cancellation, or active noise/occlusion control \cite{denk_equalization_2018, schepker_null-steering_2019,liebich_active_2016-1}. 
These approaches require knowledge, estimation or modelling of the relative transfer function from the In-Ear microphone to the eardrum, which is in the following referred to as \ac{RECTF}. Since such relative transfer functions and their properties have only been assessed very sparsely before \cite{vogl_transfer_2018}, the \ac{RECTF} is given special consideration here.

The database is made openly available. We believe that it is useful for general research and development of in-the-ear type hearing devices. This device style has gained popularity again with the advent of hearables and other hearing devices targeted at normal-hearing users in different applications \cite{rumsey_headphone_2019}. Moreover, developments based on the data can be directly transferred to portable real-time prototypes due to the joint availability and compatibility of this database, the Hearpiece \cite{denk_one-size-ts-all_2019}, and portable signal processing platforms \cite{pavlovic_open_2019, herzke_open_2017}.

\section{Methods}
\subsection{Earpiece and Measurements in the Ear}
\begin{figure}[bt]
	\centering
	\includegraphics[width = 65mm]{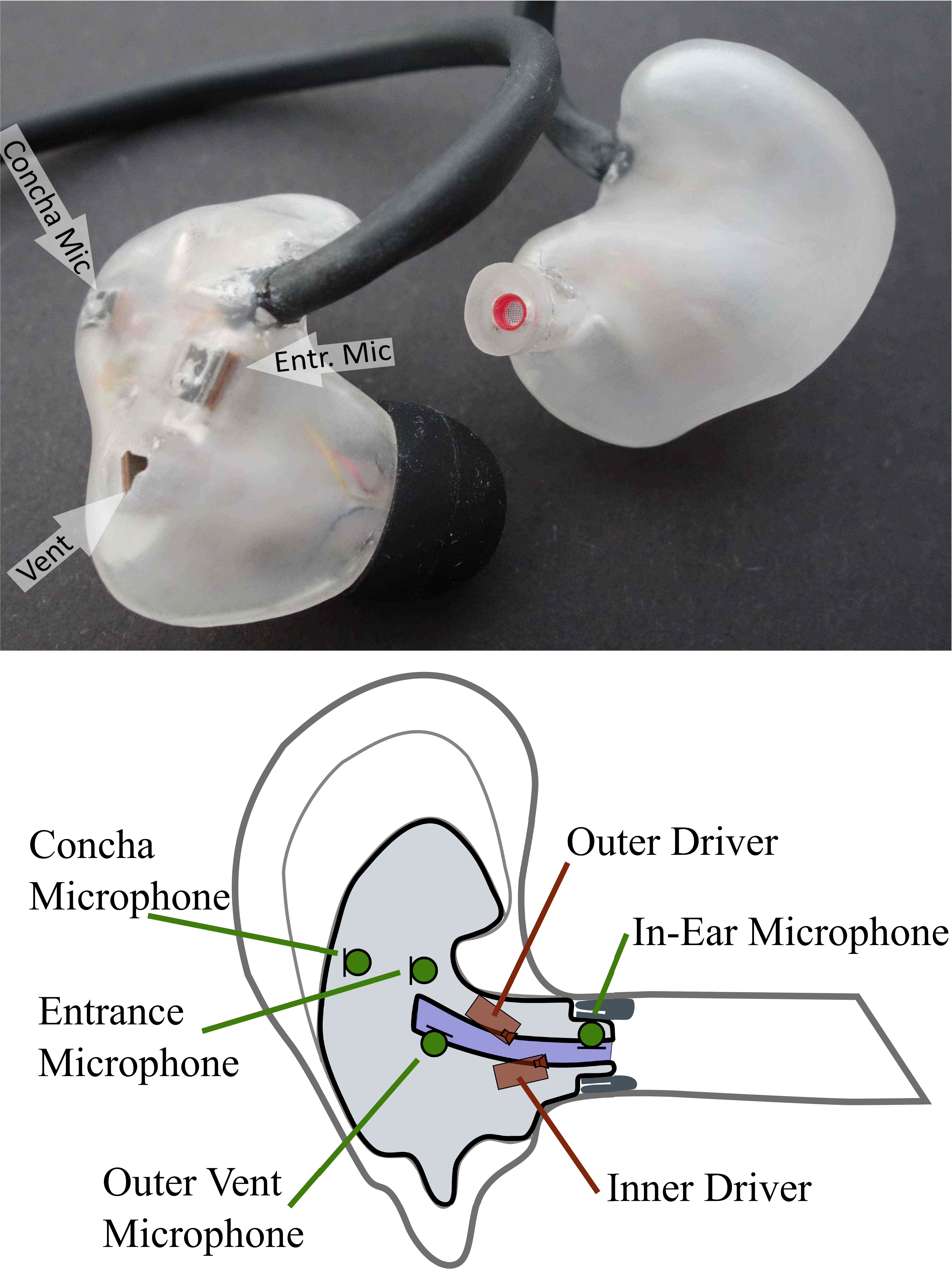}
	\caption{Top: Photograph of the Hearpiece used for recording the database. Bottom: Schematic transducer layout in the device (grey), both drivers and two microphones couple into a vent (blue area). Reproduced from \cite{denk_one-size-ts-all_2019}.}
	\label{fig:Pic-Earpiece}
\end{figure}

A photograph of the Hearpiece and its schematic layout are shown in Fig. \ref{fig:Pic-Earpiece}, the detailed geometry is documented in \cite{denk_one-size-ts-all_2019,hoertech_technical_2019}. The Hearpiece includes two Balanced-Armature drivers and four microphones in each side that are contained in an acrylic shell with a generic fit that suits about 90\% of human ears. An optimized fit is achieved using exchangeable silicone domes in four sizes, the size selected for each subject is given with the database.

Both drivers and two of the microphones are distributed along the included vent with a cross-section of approx. 1.5 mm$^2$ and a length of 19 mm, where the two microphones are positioned at the inner and outer ends (referred to as In-Ear and Outer Vent microphones, respectively). The two drivers couple into the vent at different positions and are referred to as inner and outer driver. Two more microphones are located on the faceplate with a distance of 12.5 mm, one near the position of the ear canal entrance (Entrance microphone) and one in the rear part (Concha microphone). The drivers are two different types (inner: Knowles WBFK-30042, outer: FK-26768), while all microphones are MEMS microphones of the same type (Knowles SPH1642HT5H-1). The Hearpiece  is available as both a vented and a closed version where the outer part of the vent is occluded and the Outer Vent microphone omitted. The main body of the data presented here regards the vented version; for differences of the closed version the reader is referred to \cite{denk_one-size-ts-all_2019}. The transducers were connected to the measurement system through a custom adaptor and amplifier box without implementing any real-time processing. 

In addition to the microphones integrated into the device, measurements were also conducted at the eardrum. To this end, an audiological probe tube connected to an Etymotics ER7C microphone was inserted into the ear canal until the subject reported contact with the eardrum, and then pulled back by a minimal amount. The device was then inserted on top of the probe tube. 
To minimize squeezing and movement of the probe tube, it was placed at the lower anterior corner of the ear canal, led towards the eardrum on the lower side of the ear canal and out of the ear between tragus and anti-tragus. Insertion of the probe tube and the device was executed by an experienced hearing aid acoustician.

\subsection{Individual Subjects and Dummy Head}
27 human subjects (age 24-60, average age 30.56, 13 females, 14 males) participated in the measurements. By means of screening measurements it was assured that they had clinically normal hearing (hearing threshold better than 20 dB HL for frequencies $<$ 8 kHz, normal loudness perception). Furthermore, an otoscopy was conducted directly before each session to ensure that the eardrum was visually normal and no large accumulations of cerumen were present in the ear canal. 
The subjects comprised one author, employees of the University of Oldenburg (Department of Medical Physics and Acoustics) and RWTH Aachen University (Institute of Technical Acoustics) as well as paid volunteers. All subjects signed a written informed consent, and the experiment was approved by the University of Oldenburg Ethics council. In two subjects, it was not possible to insert the Hearpiece properly, and their results were excluded.

The measurements were also conducted in a G.R.A.S. KEMAR 45BB-12 dummy head with anthropometric pinnae and low-noise ear simulators \cite{wille_iec_2016}. The anthropometric pinnae facilitated realistic fitting of the in-ear device, which was not possible with the standard pinnae. Measurements and evaluation were conducted identically for the KEMAR and the human subjects except for the eardrum data, where KEMAR's ear simulators were utilized.

\subsection{Apparatus and Procedure}
\label{sec:Procedure}
The measurements were conducted in the Virtual Reality Lab of Oldenburg University, which is an anechoic chamber with 94 Genelec 8030 loudspeakers installed in a 3D layout. 48 loudspeakers were spaced uniformly in the horizontal plane, leading to an azimuth resolution of 7.5$^\circ$. Further circles of loudspeakers were installed at $\pm30^\circ$ and $\pm60^\circ$ elevation with a horizontal resolution of 30$^\circ$ and 60$^\circ$, respectively, as well as each one loudspeaker directly above and below the center. 8 further loudspeakers were installed in the median sagittal plane to achieve a vertical resolution of 15$^\circ$ in this plane above -30$^\circ$ elevation. The seven incidence directions at elevations of $\leq-60^\circ$ could not be considered due to obstruction by a sitting platform. The loudspeakers are mounted at a distance of 2.5-3 m from the subject, and the the spatial separation of woofer and tweeter (approx. 1.3$^\circ$) can be neglected.
\begin{figure}[bt]
	\centering
	\includegraphics[width = 65mm]{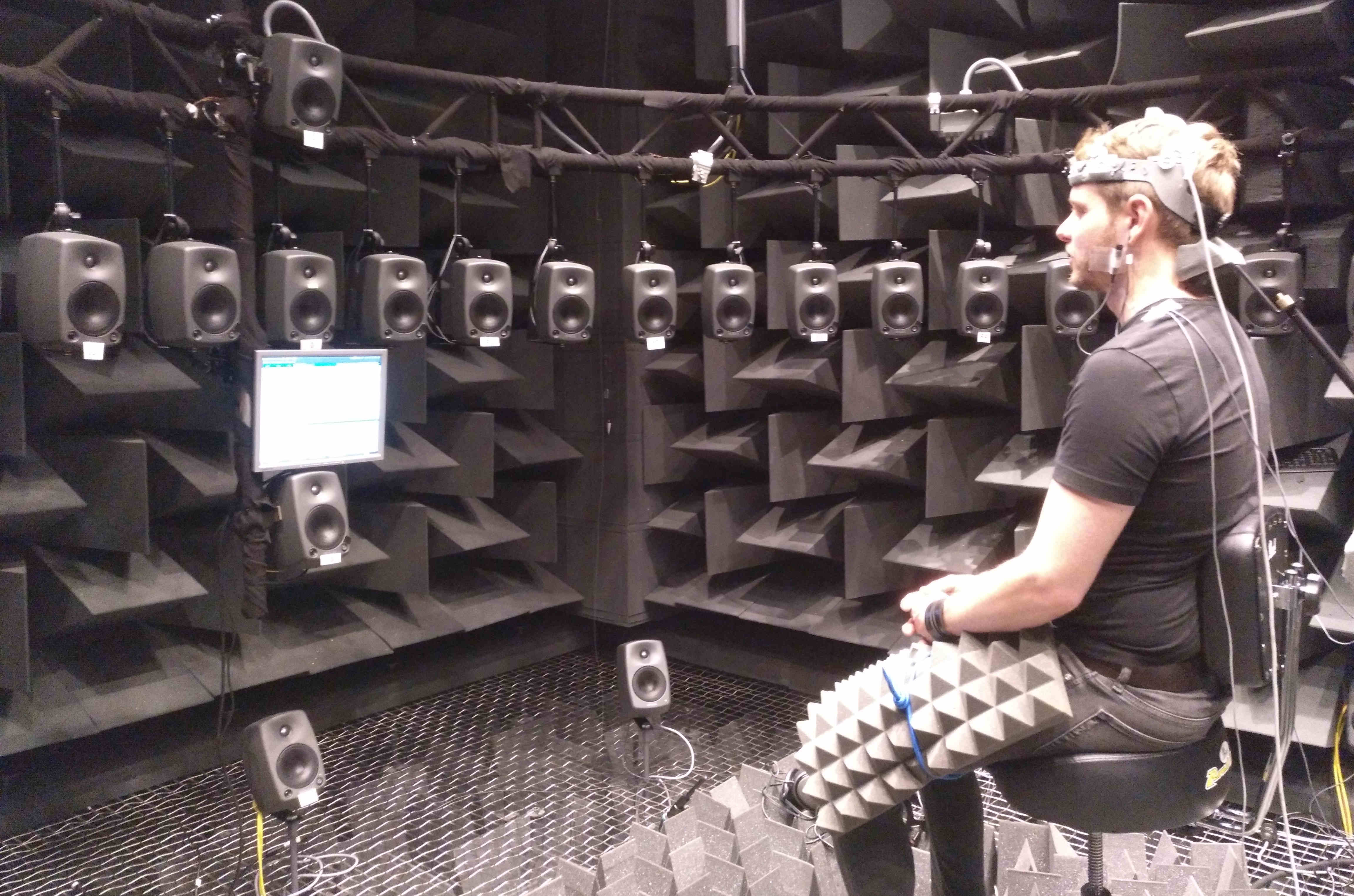}
	\caption{Experimental setup with a subject.}
	\label{fig:Photo_InTheLab}
\end{figure}

Figure \ref{fig:Photo_InTheLab} shows a photograph of the setup with a subject during an experimental session. The subjects were seated on a small grillage platform covered by absorbers, with their legs wrapped in absorbing material to minimize reflections. To stabilize the head position during the course of the experiment, a small headrest as well as a graphical feedback on the head position \cite{denk_controlling_2017} was provided. The graphical feedback utilized head position data continuously recorded using a headtracker (Pohlemus Patriot) mounted on top of the subject's head to display the corrections necessary to restore a reference head position and orientation. The graphical feedback was displayed on a screen mounted below the loudspeaker in front of the subjects (see Fig. \ref{fig:Photo_InTheLab}). At the beginning of the experiment, the subjects were positioned and oriented at the center of the loudspeaker array using crossed laser markers, and the reference head position was recorded. The head position and orientation with respect to the reference position are supplied with the database.

The experimental procedure was as follows: 
First, the probe tubes were inserted into the open ear and the subject was positioned and oriented in the center of the array. Second, the \ac{HRTF} to the eardrum of the open ear was measured (see Sec. \ref{sec:HRTF_Meas}). Third, the device was inserted and the \ac{HRTF} to all microphones of the device and the eardrum measured. The hearing aid acoustician inserted the device to minimize hazard to the subjects by pushing the probe tube closer to the eardrum and avoid squeezing of the probe tube. Fourth, the responses of the device's drivers were measured (see Sec. \ref{sec:FBP_meas}), once with nothing close to the ear and once where the subject held a telephone (Galaxy S3 mini, turned off) close to their right ear. Then, the device was taken out and steps 2-4 repeated for a total of 5 reinsertions of the device to assess variations caused by uncertainty of the fit. Finally, the \ac{HRTF} with the telephone held close to the right ear was measured both with the device inserted (no reinsertion after fifth round) and subsequently the open ear.

The database was measured with the device with serial number DV-0001, except for additional measurements that assessed nine devices with serial numbers 0003-0011. These additional measurements were gathered as described above but exclusively in KEMAR, with only three reinsertions and without the telephone nearby, about five months after the main measurements were finished.

\subsection{HRTF Measurements and Processing}
\label{sec:HRTF_Meas}
Measurements and processing of the \acp{HRTF} was performed very similar to \cite{denk_adapting_2018}. \acp{HRTF} were measured for all incidence directions where loudspeakers were installed using exponential sweeps covering a range from 30 Hz to 22.05 kHz (= half sampling rate of 44.1 kHz) with an individual length of 3.2 s. To speed up the measurements, the sweeps were overlapped in time using the multiple exponential sweeps technique \cite{majdak_multiple_2007}, leading to an overall duration of 27 s. The average level of the sweeps was 81 dB SPL in free field. For each round of measurements, the order of incidence directions was randomized independently. 

From the raw impulse responses, acoustic reflections from equipment were removed by frequency-dependent truncation \cite{denk_removing_2018}. Next, the responses were compensated for the influence of the loudspeakers (measured using 1/8" microphone G.R.A.S. 46DP-1, acoustic reflections removed likewise) and microphone sensitivities by regularized spectral division. For the microphones included in the device, one representative sensitivity (including pre-amplifiers), extended by individual broadband adaptation, was utilized. For the probe tube microphones, individual free-field sensitivities were determined. Finally, the \acp{HRTF} were set to the expected 0 dB at frequencies below 60 Hz, shifted in time by 44 samples and truncated to a final length of 356 samples at 44.1 kHz sampling rate.

\subsection{Driver Responses at the Eardrum and Feedback Paths}
\label{sec:FBP_meas}
The linear responses of all drivers were measured sequentially, and for each driver simultaneously at the eardrum and all microphones of the device. Again, an exponential sweep covering the frequency range from 30 Hz to 22.05 kHz (=half sampling rate of 44.1 kHz) with a length of 2 seconds was employed. The envelope of the sweep was pre-equalized to achieve a level of approx. 80 dB SPL equivalent to the free field at the eardrum \cite{muller_transfer-function_2001}. Afterwards, the microphone sensitivities as well as the delay and sensitivity of the sound card were compensated, such that the impulse responses are stored in units of Pa/V. Finally, the impulse responses were truncated to 756 samples at 44.1 kHz, including 44 samples time shift as in the \acp{HRTF}.

\section{Results \& Analysis}
In the following, exemplary results that aim to represent the extensive database as well as possible are presented and discussed. Explicit sample data is shown for the KEMAR and two representative human subjects: One male subject where the device fit particularly well (ER03ED10) and one female subject where only a less reliable fit could be achieved (EL08RD06). In Secs. \ref{sec:Res-HRTFs}-\ref{sec:Res-RECTF}, various acoustic parameters are assessed and shown for one sample device and insertion (DV-0001, third insertion into right ear). The variation of these parameters across human subjects is evaluated and compared to measurements in KEMAR. Averages of responses in the human subjects were computed from the magnitude responses in decibels. In Sec. \ref{sec:Res-Reinsertion}, the variability of the parameters with reinsertion of the device is evaluated for the three exemplary ears. In Sec. \ref{sec:Inter-Device}, the variation of driver responses and feedback paths within each 8 vented and 2 closed pairs of devices measured in KEMAR is shown. Finally, Sec. \ref{sec:Res-ApplEx} demonstrates a possible application scenario where real-time processing with different insertion gain settings is simulated for the device in a sample human ear.

While aspects particular to the shown data are discussed with the presentation of the Results, an overarching discussion is given in Sec. \ref{sec:DiscSum}.

\subsection{Head-Related Transfer Functions}
\label{sec:Res-HRTFs}
\begin{figure}[!b]
	\centering
	\includegraphics[width = \widthOnecol]{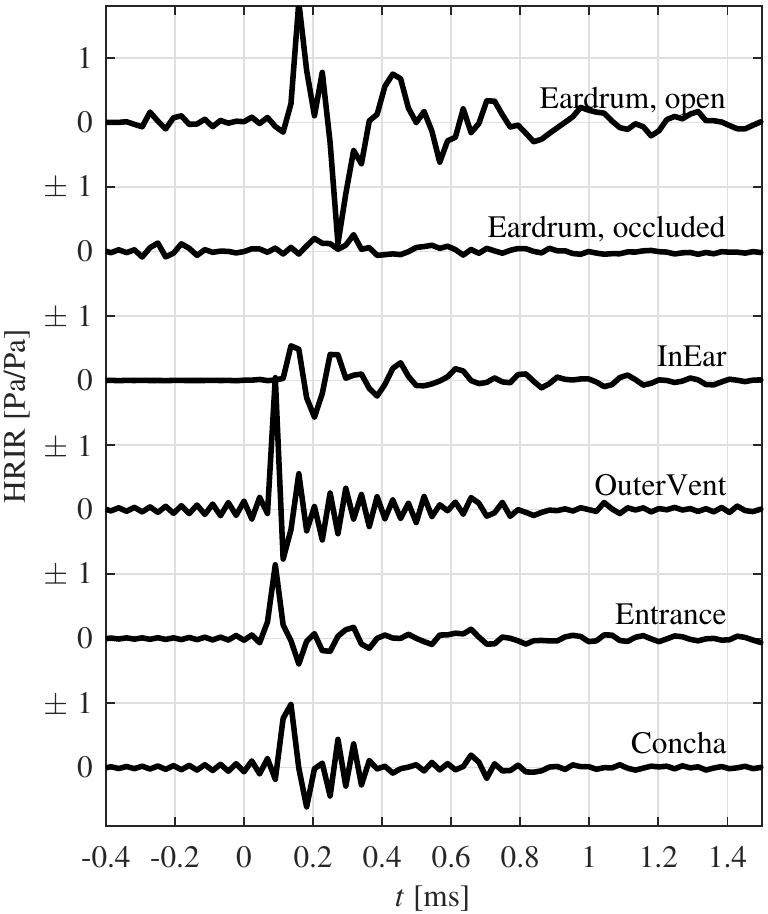}
	\caption{HRIRs for frontal incidence in the right ear of subject ER03ED10. Each line indicates the HRIR for one microphone location as indicated. The delay between channels  is caused by geometric distances.}
	\label{fig:HRIR_samples}
\end{figure}

Figure \ref{fig:HRIR_samples} shows samples of the \acp{HRIR} i.e., the impulse response representation of \acp{HRTF}, measured for frontal incidence in subject ER03ED10, for all microphone locations of interest as denoted at the right of the panel. First, small timing differences up to 0.1 ms between microphones originate from the geometric propagation difference. It should be noted that the open eardrum \ac{HRIR} was measured separately from the other responses, and the good temporal alignment between both measurements at the eardrum verifies the stability of the head position throughout the experimental session \cite{denk_controlling_2017}.
Second, the \ac{HRIR} at the eardrum of the open ear is considerably longer than the others, which is caused by an oscillation at the $\lambda/4$ resonance frequency of the ear canal. The level differences between microphone locations are caused by attenuation through insertion of the device (Eardrum, occluded) or due to (destruction of) such resonances of the open ear \cite{denk_adapting_2018}.
Third, in the \acp{HRIR} measured at the eardrum, additional acausal peaks are seen at around -0.3 ms in both curves, but not in the \acp{HRIR} measured in the microphones of the device or those of the KEMAR. These peaks are not to be confused with the mild pre-ringing artefacts as present in the Outer Vent or Concha microphones. The acausal peaks very likely originate from a sound path leaking directly into the body of the probe tube microphone without travelling through the tube. This interpretation is consistent with the additional observation that the temporal alignment of this component with respect to the main response varies with incidence direction. As it can be seen in Fig. \ref{fig:HRIR_samples}, the acausal parts are most critical in the occluded eardrum data, due to the lower level of the main response that is attenuated with respect to the open-ear case. In the occluded eardrum responses, the additional energy of this disturbing component may impose a lower boundary on the measured \acp{HRTF}. Nevertheless, comparisons to KEMAR data (see following sections) and further analyses showed that up to about 10 kHz, the \acp{HRTF} measured at the occluded eardrum yield results that are as reliable as it can be expected from probe tube measurements in the occluded ear. In the responses at the eardrum of the open ear, this disturbance is generally low enough in level to be neglected.

\begin{figure}[bt]
	\centering
	\includegraphics[width = \widthOnecol]{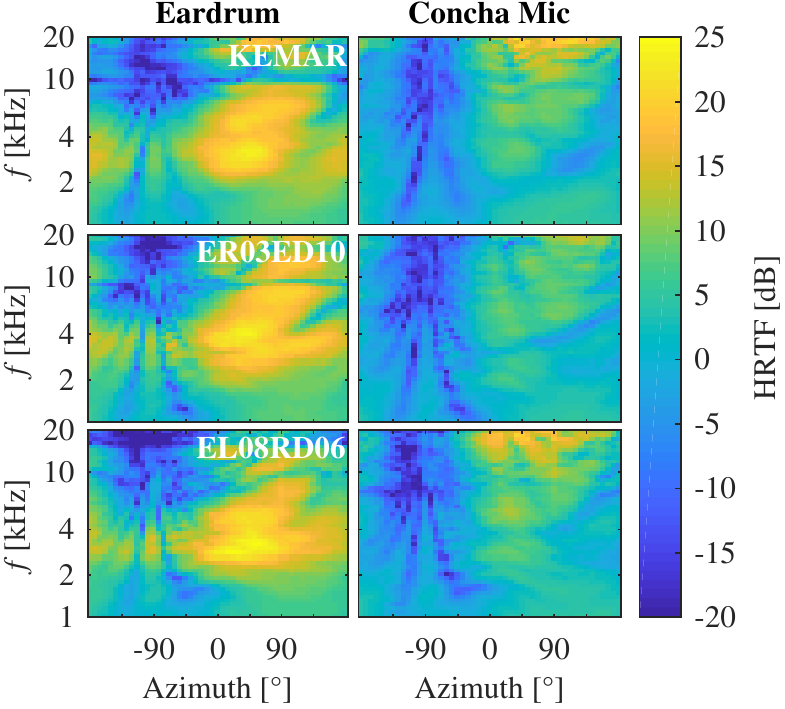}
	\caption{HRTFs in the horizontal plane measured in KEMAR and two representative human subjects, at the eardrum of the open ear (left panels) and the Concha microphone of the device (right panels).}
	\label{fig:HRTF_samples}
\end{figure}

Figure \ref{fig:HRTF_samples} shows a direction-frequency representation of the \ac{HRTF} at the eardrum of the open ear and the Concha microphone after 1/12 octave smoothing, for the three exemplary subjects. Somewhat altered but similar structures are seen in all subjects for each microphone location. As expected, the \acp{HRTF} differ notably between microphone locations in each subject. The most prominent difference is an amplification around 2-10 kHz originating from ear cavity resonances, which is seen at the eardrum but not at the Concha Microphone \cite{denk_adapting_2018,shaw_sound_1968}. While the differences between left and right hemispheres are evident at both microphone locations, the eardrum \ac{HRTF} also contains more spatial dependences than the Concha Microphone that originate from directional pinna filtering, since the pinna is largely filled by the device \cite{denk_spectral_2018, durin_acoustic_2014}.

\subsection{Insertion Loss}
\label{sec:Res-InsLoss}
\begin{figure}[bt]
	\centering
	\includegraphics[width = \widthOnecol]{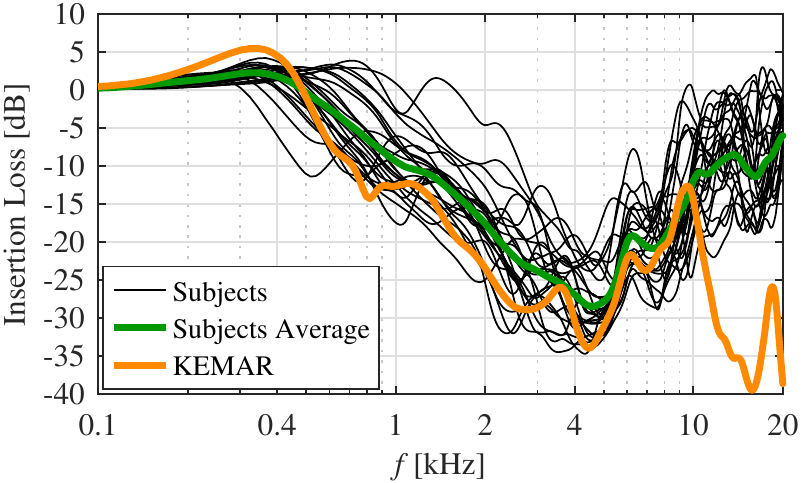}
	\caption{Insertion Loss for diffuse-field incidence measured in individual subjects (black curves), the average across subjects (green curve) and in KEMAR (orange curves).}
	\label{fig:InsertionLoss}
\end{figure}

Figure \ref{fig:InsertionLoss} shows the diffuse-field insertion loss, i.e., the attenuation of external sounds reaching the eardrum by inserting the passive device. The insertion loss was calculated by dividing the approximated diffuse-field responses at the occluded eardrum by the appropriate open-eardrum response. The diffuse-field responses were approximated from the \acp{HRTF} by calculating the power spectrum average across 47 uniformly distributed incidence directions after 1/12 octave smoothing of individual \acp{HRTF} \cite{denk_adapting_2018}. Each black line in Fig. \ref{fig:InsertionLoss} denotes the result for one right ear of a human subject in the third insertion, the green line denotes the average across subjects, and the orange line denotes the appropriate result in KEMAR.

The typical insertion loss curve approaches 0 dB at the low frequencies, i.e., the vent allows low-frequency sounds to leak into the ear canal unattenuated. Around roughly 400 Hz, an amplification of up to 5 dB is seen most prominently in the KEMAR data, but also for some of the human subjects. The amplification probably results from a Helmholtz resonance of the residual ear volume and vent opening. Above approx. 500 Hz, the attenuation increases for most subjects up to about 30 dB around 4 kHz. Only in some subjects, where only a poor fit could be achieved, the device does not attenuate sounds below 1-2 kHz. It should be noted that even poorer fits than included here occur in the database, and within some subjects the fit varies significantly between insertions (see also Sec. \ref{sec:Res-Reinsertion}). Between 4 and 10 kHz, the attenuation decreases again down to 10 dB, which might be caused by approaching the $\lambda/2$ resonance of the vent (length: 19 mm). Above 10 kHz, the insertion loss increases again for KEMAR measurements, but decreases further in the human subjects. Apart from outliers with a very poor fit, the insertion loss in the human subjects lies within a range of approx $\pm$7 dB around the average for frequencies $>$600 Hz.

Up to 10 kHz, the data from the human subjects and KEMAR are very consistent, and the KEMAR may be seen as a human subject where a particularly good fit could be achieved. In the human subjects, the presence of the probe tube unavoidably prevents a tight seal between the ear canal and the silicone dome of the earpiece. Thus, in practice the low-frequency insertion loss in users may look even more like the KEMAR curve.
Above 10 kHz, the results deviate consistently between human subjects and KEMAR. The lower attenuation seen in the human subjects could, on the one hand, be caused by utilizing KEMAR out of its intended frequency range \cite{burkhard_anthropometric_1975}. On the other hand, we believe it is more likely that the apparently lower attenuation in the human subjects is an artefact of a low \ac{SNR} in this frequency region of the occluded eardrum measurements (c.f. Sec. \ref{sec:Res-HRTFs}).

\subsection{Driver Responses at Eardrum}
\label{sec:Res-HpTFs}
\begin{figure}[bt]
	\centering
	\includegraphics[width = \widthOnecol]{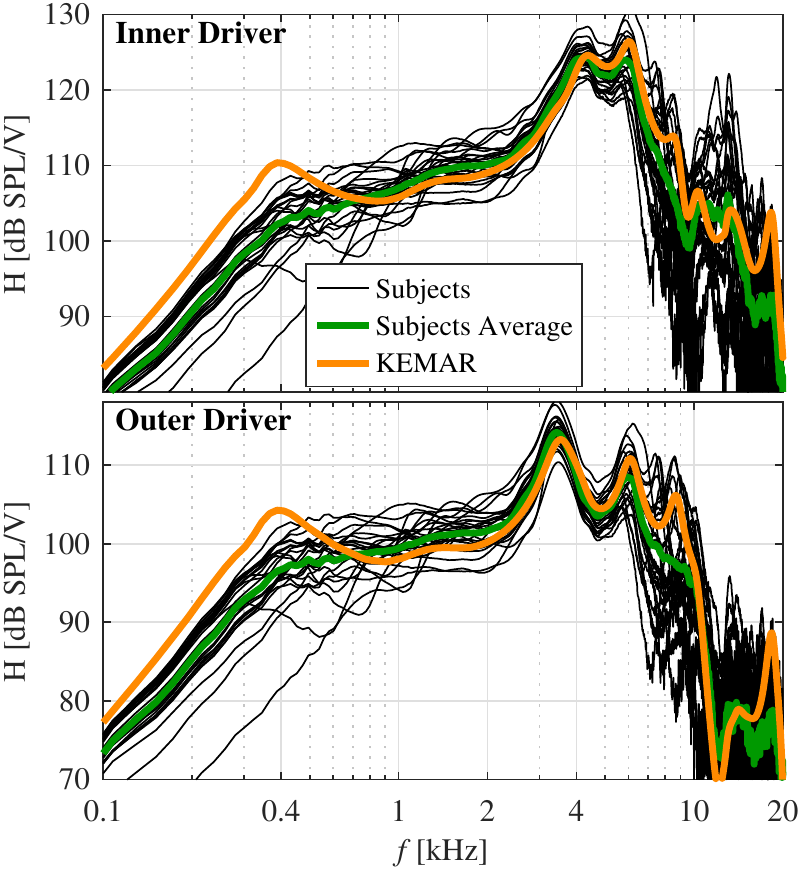}
	\caption{Responses of the inner (upper panel) and outer (lower panel) drivers of the device measured at eardrum. Black lines show the responses measured in individual subjects, thick orange line measured in KEMAR (left and right ears plotted) in both cases.}
	\label{fig:DrResp}
\end{figure}

Figure \ref{fig:DrResp} shows the responses of both drivers of the device at the eardrum in separate panels. Generally, the responses between drivers differ, which is caused by using different driver types \cite{denk_one-size-ts-all_2019}. A low-frequency roll-off with cut-off frequencies varying between approx. 300 Hz and 1 kHz is seen in all curves is caused by incomplete sealing of the ear canal due to the vent and imperfect fit. As for the insertion loss, this inter-subject difference in this frequency range is probably caused by fits with varying tightness in the ears of subjects. The tight seal that can be achieved in the KEMAR also here leads to the lowest cut-off at around 300 Hz; for most subjects it lies at around 400 Hz. In the KEMAR data, a resonance around 400 Hz can be seen that is also observed less prominently in the human data (see also Sec. \ref{sec:Res-Reinsertion}).

In the range of 2-6 kHz, the variation between subjects is small ($< \pm 5$\,dB) and mostly comprises a broadband offset. This can be explained by the fact that the corresponding half wave lengths are still larger than the residual ear canal lengths, and the responses are largely governed by the properties of the drivers and the device. The broadband differences are probably a result of residual ear canal volumes that differ between subjects. Beyond 6 kHz, the resonances in the individual residual ear canals come into play, which lead to shifted resonances and between-subject deviations of 30 dB and more. 

The KEMAR data reflects a median human response well across the full frequency range, and is very close to the human average above 800 Hz. As for the insertion loss (Sec. \ref{sec:Res-InsLoss}), it can be assumed that in human subjects a better seal than in the presented data can be achieved in practice due to the absence of the probe tube. Therefore, the actual low-frequency response in human subjects may be even closer to the KEMAR data than in the present results.

\subsection{Feedback Paths}
\label{sec:Res-FBPs}
\begin{figure}[bt]
	\centering
	\includegraphics[width = \widthOnecol]{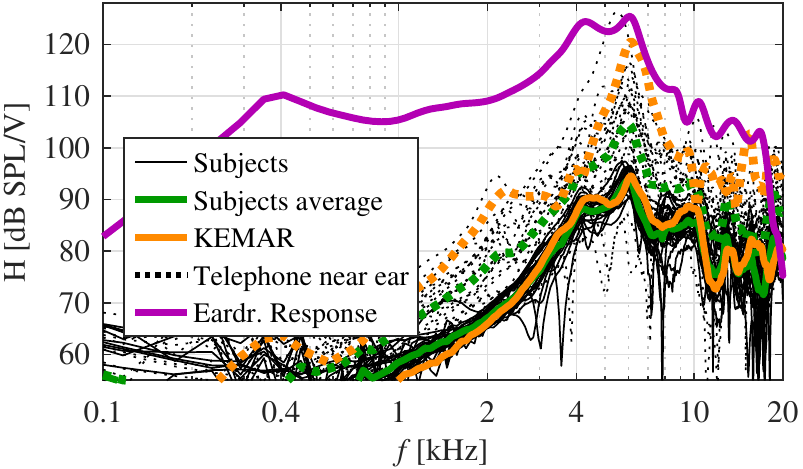}
	\caption{Feedback paths from the inner driver to the Concha microphone in individual subjects (black lines) and the KEMAR (orange line), response at eardrum for reference (purple line, measured in KEMAR).}
	\label{fig:FBPaths_subj}
\end{figure}
Figure \ref{fig:FBPaths_subj} shows feedback paths, i.e., the responses of the inner driver at the Concha microphone. The general behaviour for paths including either driver and the Concha and Entrance microphones, which are most relevant for incoming signal pickup in most applications, are very similar \cite{denk_one-size-ts-all_2019}. Free-field conditions are denoted by solid lines, conditions when a telephone was held near the right ear are shown as dotted lines. For either condition, again the individual subject are denoted by black lines, the subject average is denoted by green lines and the KEMAR result is denoted by orange lines in the appropriate line style. For reference, the response of the inner driver at the eardrum measured in KEMAR is also shown as a purple line.

The feedback path in free-field conditions is generally lower in level than the response at the eardrum by more than 30 dB up to 4 kHz, and about 20 dB at higher frequencies. The lower boundary of the feedback paths at around 60 dB SPL/V (around 50 dB SPL/V in the KEMAR) is probably caused by noise in the measurements due to the level limitations. It is expected that below 2 kHz, the downward slope of approx. 24 dB per octave observed between 2 and 4 kHz actually continues toward lower frequencies. Placing a telephone near the ear generally results in a rather broadband amplification of the feedback paths of approx. 10 dB in average. However, the influence of the telephone is subject to large variations between subjects, presumably due to different manners of how they held the telephone. 

The variation across subjects for the free-field case is in the range of the between-subject variation of the driver responses at the eardrum (c.f. Sec. \ref{sec:Res-HpTFs}). The KEMAR results lie well in the range of human subjects data, and coincides almost perfectly with the human average curve. In the condition with the telephone near the ear, the increase of the feedback path in KEMAR is rather high but still in the range of human data, presumably because the telephone was placed very close to the ear.

\subsection{Residual Ear Canal Transfer Functions}
\label{sec:Res-RECTF}
\begin{figure}[bt]
	\centering
	\includegraphics[width = \widthOnecol]{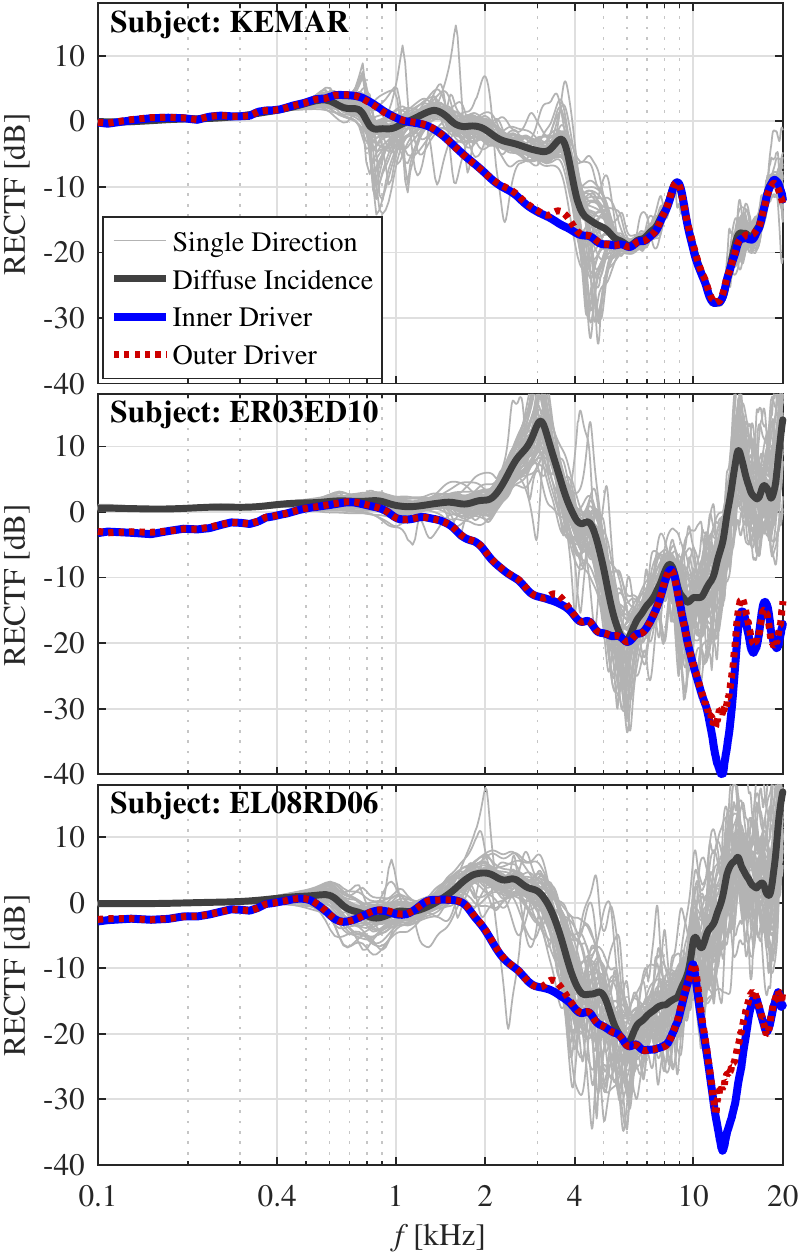}
	\caption{Residual ear canal transfer function (RECTF, from In-Ear microphone to eardrum) measured in KEMAR and two representative human subjects. Results are shown for different sound sources as denoted by the legend.}
	\label{fig:RETF_samples}
\end{figure}
Figure \ref{fig:RETF_samples} shows the \ac{RECTF} (the relative transfer function between the In-Ear microphone and occluded eardrum) in the three representative ears for external sound sources as well as the two drivers. The \ac{RECTF} was computed by dividing the appropriate 1/12 octave smoothed magnitude responses at both locations. It can generally be seen that the \ac{RECTF} is only a flat 0 dB in the low frequencies below 400 Hz, and generally has to be considered when estimating the sound pressure at the eardrum using the In-Ear microphone \cite{vogl_individualized_2019}. The sign of the \ac{RECTF} shows that the sound pressure level is mostly higher at the In-Ear microphone, at some frequencies this difference amounts to 30 dB and more.

Besides the dependence on frequency, considerable differences between sound sources are noted. That is, the \ac{RECTF} is different between external sound sources and the two included drivers of the device. While the \ac{RECTF} differs between external sound sources at different directions, it is very similar between the two drivers as sound source. This deviation between sound sources is most prominent in a band between approx. 1.5 and 5 kHz; in the human subjects it is also seen above 10 kHz. The results are consistent with observations made in a previous prototype of the Hearpiece \cite{vogl_transfer_2018}, however the underlying reason is still unclear and not within the scope of the present paper.

\begin{figure}[bt]
	\centering
	\includegraphics[width = \widthOnecol]{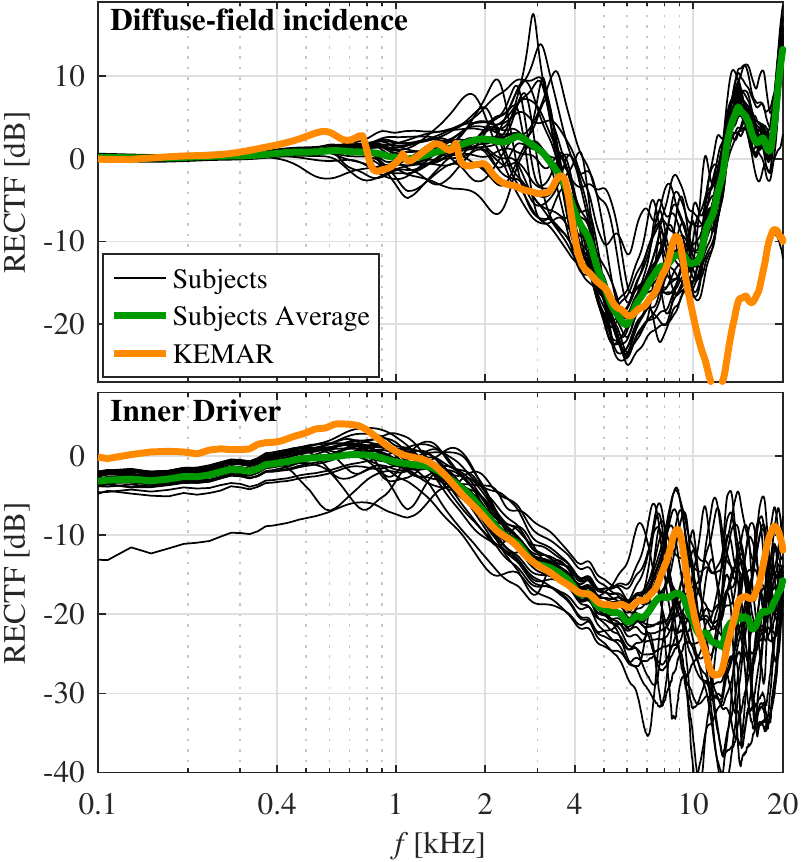}
	\caption{Residual ear canal transfer function (RECTF, from In-Ear microphone to eardrum) measured in all subjects and the KEMAR, for diffuse-field incidence (upper panel) and the inner driver of the device (lower panel). The results for the outer driver are very similar to the inner driver (c.f. Fig. \ref{fig:RETF_samples}).}
	\label{fig:RETF_average}
\end{figure}

The \ac{RECTF} measured in KEMAR in relation to the data from all human subjects is shown in Fig. \ref{fig:RETF_average} for diffuse-field incidence and the inner driver. The KEMAR data is very consistent with the human data up to 10 kHz for diffuse-field incidence, and across the whole frequency range for the drivers. The deviations between the KEMAR and human data for diffuse-field incidence across 10 kHz might again be explained by a low \ac{SNR} at the probe tube microphone in this measurement (see Sec. \ref{sec:Res-HRTFs}). The hypothesis is supported by the fact that the between-subject differences are decreasing with increasing frequency in this range for diffuse-field incidence, but increasing for the inner driver as a sound source. Between-subject differences that are increasing with frequency would generally be expected due to increasing differences between ear canal geometries relevant in this frequency range.

\subsection{Reinsertion Variability}
\label{sec:Res-Reinsertion}
\begin{figure}
	\centering
	\includegraphics[width = \widthOnecol]{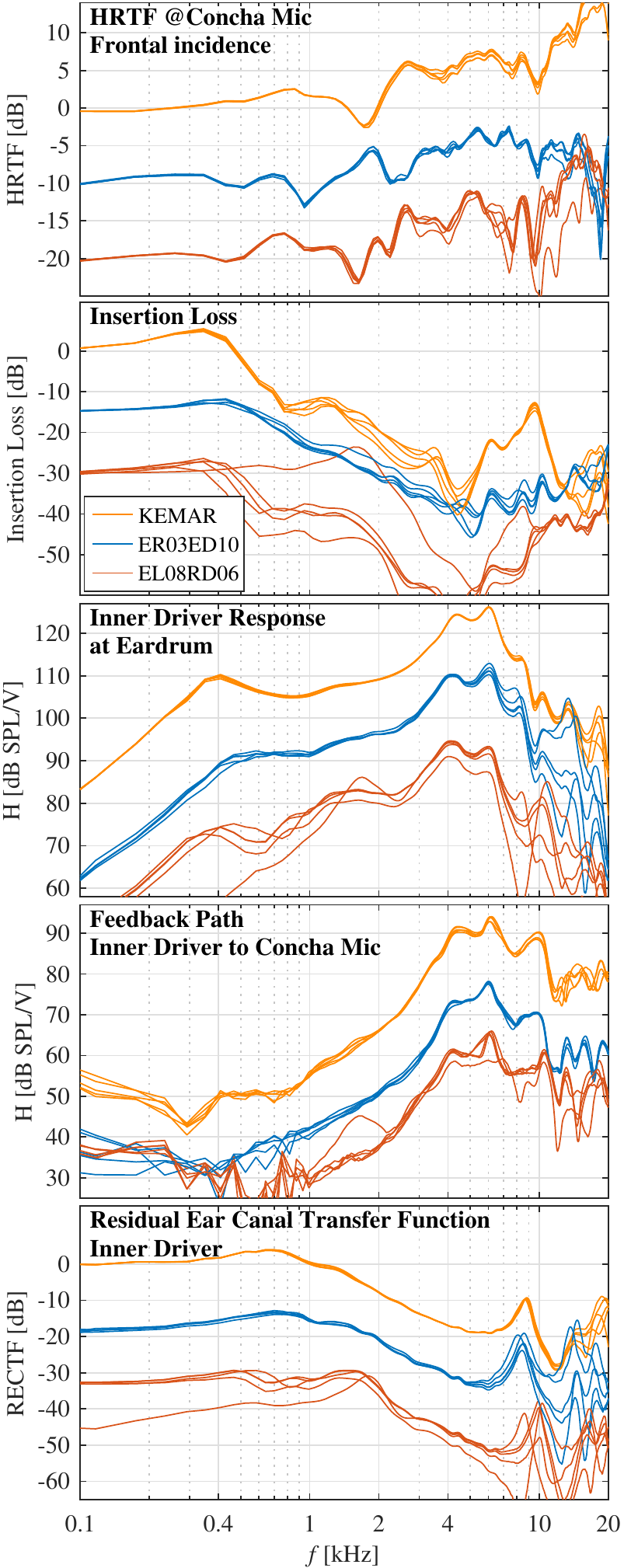}
	\caption{Repositioning accuracies for the three sample subjects, each line showing data for one insertion of the device in the left ear. Each panel shows a different quantity as denoted in the panel title, data for the individual subjects have been shifted for better display.}
	\label{fig:Repositioning}
\end{figure}
Besides the variation between ears, the transfer functions may change due to movements in the ear, or when the devices is reinserted. Figure \ref{fig:Repositioning} shows the the transfer functions that were assessed before for all five measured insertions into the three exemplary ears. 
While some inter-subject differences are depicted well in this Figure, in this section we only discuss particular effects of reinsertions. It should be stated here again that the device was inserted not by the subjects themselves but by the hearing aid acoustician (c.f. Sec. \ref{sec:Procedure}).

The top panel of Fig. \ref{fig:Repositioning} shows the \ac{HRTF} measured at the Concha microphone for frontal incidence. Hardly any reinsertion variation is seen, even at frequencies approaching 20 kHz. The exception is one insertion in subject EL08RD10 where a poor fit was achieved, which is discussed in more detail below. The results show that the \ac{HRTF} is rather stable. On the one hand, the fit in the cavum concha seems to be well reproducible for most ears. On the other hand, many directional effects of pinna diffraction observed in the open ear are already reduced at the Concha microphone (see Fig. \ref{fig:HRTF_samples}), leading to generally smooth \acp{HRTF}.

The Insertion Loss is shown in the second panel of Fig. \ref{fig:Repositioning}. The effect of repositioning is a bit stronger than for the \ac{HRTF}, but lies within $\pm3$\,dB in each subject with the exception of one poor insertion in subject EL08RD06. It is expected that the good reinsertion accuracy seen in the controlled lab conditions can also be approximated in practice at least for the insertion loss, since subjects would be able to sense an occurring leak rather well.

The effect of repositioning on the responses of the inner driver is shown on the third panel of Fig. \ref{fig:Repositioning}. Below 8\,kHz, only a very small effect of reinsertion $<\pm2$\,dB is seen. The exception is again one insertion in subject EL08RD06, where a poor fit resulted in both a reduced insertion loss, and in the driver response to a higher high-pass cut-off frequency and an altered high-frequency behaviour. A considerable reinsertion variation of the driver response above 10 kHz is seen in the human subjects but less pronounced in KEMAR, which may (partly) be caused by unintended small movements of the probe tube rather than an actual variation of the driver response.

The effect of repositioning on the feedback paths (inner driver to Concha microphone) is shown in the fourth panel of Fig. \ref{fig:Repositioning}. A generally very low reinsertion variability $<\pm2$\,dB is seen in a frequency range between 1 kHz (sufficient SNR, c.f. Sec. \ref{sec:Res-FBPs}) and 10 kHz, equally in the subjects and the KEMAR. The exception is again the data from the insertion in subject EL08RD06 where only a poor fit was achieved. However, it should be noted that the change in the feedback path due to poor fitting in this subject is much smaller as compared to the effect on the insertion loss or driver response. Even beyond 10 kHz, the reinsertion accuracy is in the range of $\pm5$\,dB.

Finally, the bottom panel of Fig. \ref{fig:Repositioning} shows the influence of repositioning on the \ac{RECTF} for the inner driver. Up to 10 kHz, only very small variations with reinsertion ($<\pm1$\,dB) are noted. The exception is again the single poor insertion of the device in subject EL08RD06, which causes large deviations from the other measurements, similarly to the driver response. At high frequencies, reinsertion variations similar to those seen in the driver response are noted, especially in subject ER03ED10.

\subsection{Inter-Device Variability}
\label{sec:Inter-Device}
\begin{figure}[bt]
	\centering
	\includegraphics[width = \widthOnecol]{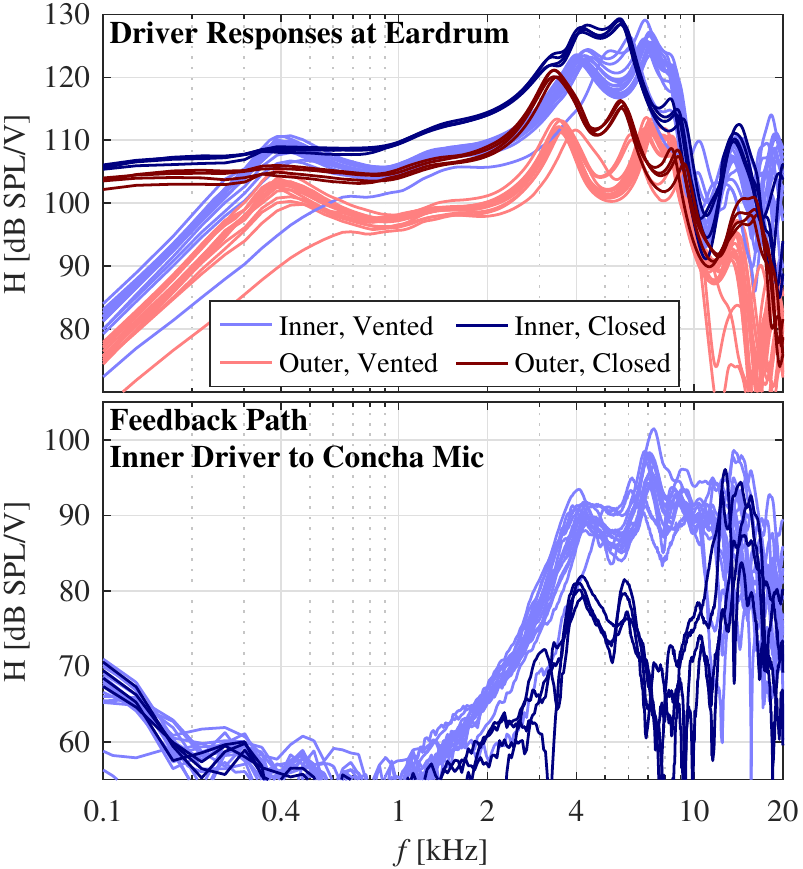}
	\caption{Serial variation of the driver responses measured in KEMAR with 8 device pairs, each line representing one device side. The inner driver results are plotted in blue, outer driver results in red. Light colors indicate vented versions of the device, dark colors the closed version.}
	\label{fig:SerialDiff_Driver}
\end{figure}
The top panel of Fig. \ref{fig:SerialDiff_Driver} shows the inter-device variation for the driver responses of a series of 10 device samples measured in KEMAR. The sample included 8 pairs of vented devices (light lines), as well as two pairs of closed devices (darker lines), both ears shown for the second insertion. The responses are very similar between devices, and the deviation can mostly be described as a broadband sensitivity offset of approx. $\pm2$ dB, except for one vented device where only a poor fit was achieved. The general differences between both drivers, as well as between open and closed devices are intended by design, further analyses regarding these differences are provided in \cite{denk_one-size-ts-all_2019}.

The between-device differences of the feedback paths (inner driver to Concha microphone, as in Fig. \ref{fig:FBPaths_subj}) are shown in the bottom panel of Fig. \ref{fig:SerialDiff_Driver}. Again, only small differences exist between devices (except between open and closed design) that are mostly in the range of the variation of the driver response.

\subsection{Application Example}
\label{sec:Res-ApplEx}
\begin{figure}
	\centering
	\includegraphics[width = 65mm] {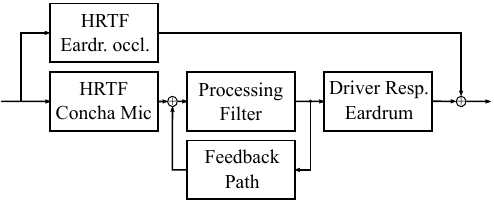}
	\caption{Flowchart for the simulation of real-time sound processing.}
	\label{fig:Flowchart_sim}
\end{figure}

\begin{figure}[tb]
	\centering
	\includegraphics[width = \widthOnecol]{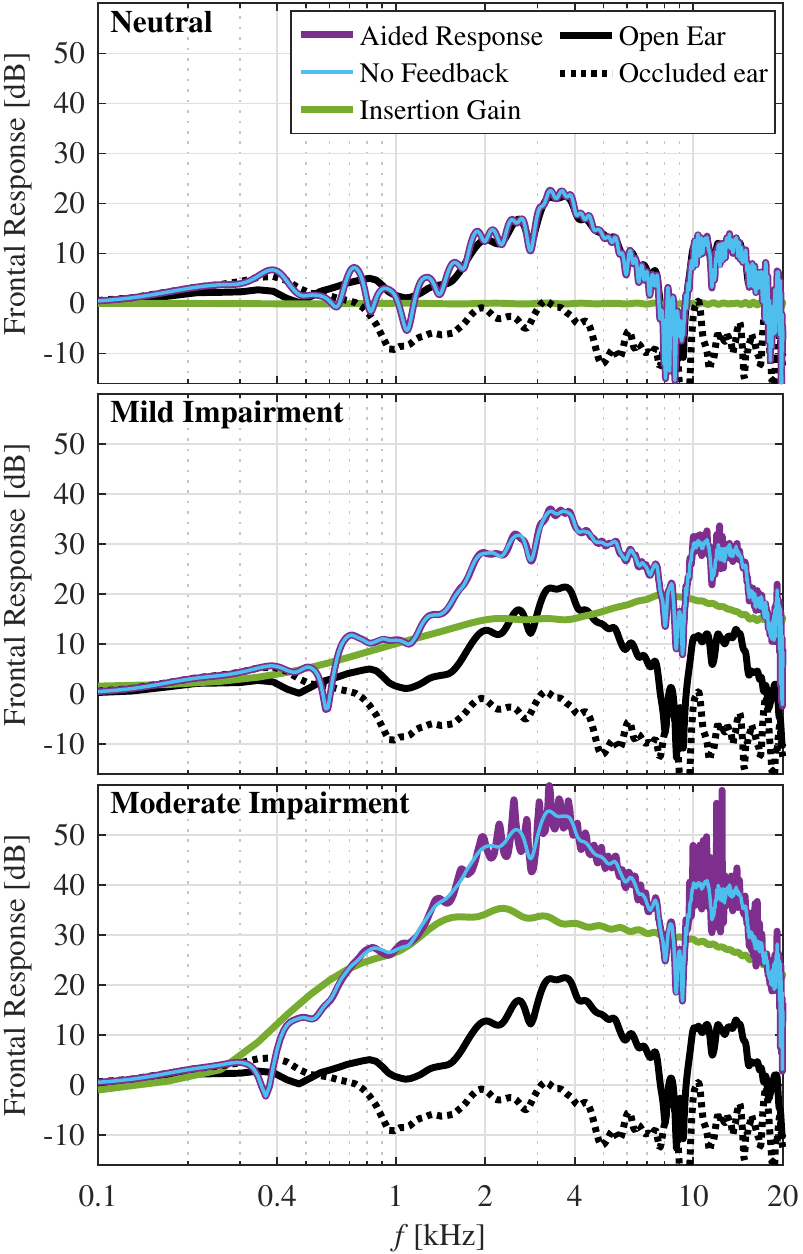}
	\caption{Obtained responses at eardrum, with simulation of real-time hearing device (purple lines) adjusted to provide insertion gains (green lines, appropriate for application denoted in panel title), open-ear (solid black lines) and occluded (dotted black lines) responses at eardrum for reference. In the No Feedback results (blue lines), the feedback path was set to 0 for calculations of the aided responses.}
	\label{fig:FullProcessing}
\end{figure}

Finally, an application example of the database is given. Real-time processing in a linear hearing device based on the Hearpiece including all sound paths to the eardrum (occluded response, device output including processing delay, feedback) was simulated for frontal incidence as depicted in Fig. \ref{fig:Flowchart_sim}. The transfer functions obtained in the right ear for subject ER03ED10 with the third insertion of the vented device were utilized. Only the Concha microphone and the inner driver were included for sound pickup and reproduction, respectively. Processing included a constant filter that was designed similar to \cite{denk_equalization_2018} such that three different insertion gain curves as denoted in Fig. \ref{fig:FullProcessing} were provided. The insertion gains were chosen arbitrarily, but could be prescribed for a neutral setting (no amplification, often referred as hear-through \cite{valimaki_assisted_2015, hoffmann_insert_2013}), a mild-sloping and a moderate-sloping hearing loss \cite{dillon_hearing_2012}. A frequency-independent processing delay of 3.5 ms was assumed for the simulations, which was attributed to the driver response at eardrum as well as the feedback paths. The influence of feedback is examined by including it in the simulations ("Aided Response" in Fig. \ref{fig:FullProcessing}) or setting it to 0 ("No Feedback"). 

The top panel if Fig. \ref{fig:FullProcessing} represents a flat 0 dB insertion gain, i.e., a neutral "hear-through" setting that would let the user hear the environment similar to the open ear \cite{valimaki_assisted_2015, hoffmann_insert_2013}. The resulting aided response is very close to the open-ear response, except for a spectral ripple below approx. 2 kHz, which originates from interferences of the occluded response with the delayed device output (c.f. \cite{denk_equalization_2018}). There is no influence of omitting feedback when calculating the aided response, i.e., the influence of feedback is negligible in this gain setting.

The middle and bottom panels show aided responses including linear amplification. The aided responses correspond well to the open-ear response with added insertion gains. Particularly with the gain setting in the bottom panel, setting the feedback paths to zero changes the simulated aided response significantly. The feedback leads to a very rippled response around the peaks of the response, and further analyses verify that this gain setting is at the upper end of the stability region. In conclusion, the vented device can be used to provide insertions gains targeted at listeners with a mild to moderate hearing loss (at least for free-field conditions). With gain settings as shown for the moderate impairment case and above, feedback management has to be included to avoid instabilities, also considering the possible level increase of feedback paths when an object is placed close to the ear (c.f. Fig. \ref{fig:FBPaths_subj}). 

\section{Discussion and Conclusion}
\label{sec:DiscSum}
We presented and evaluated a database of transfer functions of the Hearpiece, an openly available earpiece for hearing device research  \cite{denk_one-size-ts-all_2019}. The database comprises \acp{HRTF} from 87 incidence directions as well as responses of the device's drivers, all measured at the 4 integrated microphones and the eardrum in the open and occluded ear. Measurements were made in both ears of 25 human subjects and a KEMAR dummy head with anthropometric pinnae, for each five reinsertions of one sample device, as well as KEMAR measurements of 9 additional devices from a series. In total, the database amounts to 169,878 \acp{HRTF} and 5,740 driver responses/feedback paths.

The data can be used in research and development of all sorts of hearing device algorithms and applications such as hearing aids, directional microphones, noise reduction, feedback management, augmented reality or active hearing protection. A special benefit is that the algorithms can be easily transferred to portable real-time prototypes due to the joint open availability of the Hearpiece and compatible mobile processing platforms \cite{pavlovic_open_2019}. Also, influences of variations of the transfer functions on the algorithm performance can be studied with the present data. As an example, we showed results of simulating a real-time hearing device in a subject's ear with various linear gain settings. This allowed to demonstrate realistic artefacts occurring due to the venting, processing delays, and feedback (v.f. Fig. \ref{fig:FullProcessing}), and estimating the limits of achievable insertion gains. 

An integrated microphone in the ear canal as in the Hearpiece may become more common in future hearing devices. The present database provides a starting point for development of algorithms exploiting such a microphone, including individual equalization \cite{denk_equalization_2018}, feedback cancellation \cite{schepker_null-steering_2019}, or active occlusion/noise cancellation \cite{liebich_active_2016-1}. The relative transfer function between the in-ear microphone and the eardrum (here referred to as \ac{RECTF}), which has to be known for such applications, has been analysed thoroughly in this work. It was shown that this relative transfer function depends not only on frequency, the individual ear and the fit of the device. Is also differs significantly between sound sources, i.e., between the device's drivers and external sound sources, and less pronounced also between external sound sources at different positions (c.f. Figs. \ref{fig:RETF_samples} and \ref{fig:RETF_average}). Although observed before in a custom prototype similar to the present device \cite{vogl_transfer_2018, denk_individualised_2018}, the origin of this dependence is unclear and should be examined in future work. Electro-acoustic models of the device that can individually predict the transfer functions to the eardrum are under current investigation \cite{vogl_individualized_2019, roden_using_2020}

Furthermore, the database allowed an assessment of variations of the involved transfer functions associated with a variation of the fit, as well as between ears. Generally, it became obvious that these sources cause different kinds of variations. Variations of the fit cause variations of the transfer functions across the whole frequency range, most pronounced in the low frequencies below 1 kHz. A poor fit may introduce a significant leak between the device and the skin, and thus jointly leads to a smaller insertion loss (c.f. Fig. \ref{fig:InsertionLoss}), a poorer low-frequency reproduction with the balanced armature drivers (c.f. Fig. \ref{fig:DrResp}), and variation of all other transfer functions (see subject EL08RD06 in Fig. \ref{fig:Repositioning}). A poor fit acts similarly as a vent, and comes in addition to the effect of a vent that is already included in the Hearpiece. 

Repositioning also led to shifted resonances in the high frequencies $>$ 8 kHz (Fig. \ref{fig:Repositioning}, especially in the driver responses and \ac{RECTF})---however, it cannot be ruled out that this observation is an artefact of the probe tube being moved slightly during reinsertion of the device. The transfer functions that were least influenced by a variation of the fit are the \ac{HRTF} to the microphones on the faceplate, as well as the feedback paths (c.f. Fig. \ref{fig:Repositioning}). Even if a similar fit is achieved upon reinsertion or conserved during use, small variations of all transfer functions in the order of $\pm$3 dB up to 10 kHz and larger differences in the high-frequency end seem unavoidable. Similar variations jointly have to occur in timing and phase aspects of the transfer functions, which have not been explicitly assessed in the present work. Such variations of transfer functions due to variations of the fit have to be considered in the design of algorithms. 

Between subjects, differences that come in addition to variations due to the fit were observed, although it is clear that these causes cannot always be separated. Significant differences between ears were observed in all assessed metrics and in a broad frequency range. Even at low frequencies, the insertion loss and driver responses (c.f. Figs. \ref{fig:InsertionLoss}, \ref{fig:DrResp} and \ref{fig:Repositioning}) differ between subjects in a way that is different from repositioning variations. In the driver responses, feedback paths and the \ac{RECTF}, differences between subjects include broadband offsets across the whole frequency range, as well as shifted resonances in a frequency region above 8 kHz (c.f. Figs. \ref{fig:DrResp} and \ref{fig:Repositioning}). These effects mainly depend on the residual ear canal, where probably differences in the load volume lead to the observed broadband offsets, whereas the exact geometric parameters such as the length only come into play at very high frequencies. Overall, the analyses of between-subject differences show that it is worthwhile to adapt hearing devices to the acoustics of the individual ears by measuring or estimating individual transfer functions in-situ and regard them in parameter settings of algorithms.

Measurements in KEMAR with anthropometric pinnae represent a median ear very well for all assessed metrics up to 8 kHz, and reasonably well also at higher frequencies. Remaining low-frequency differences can probably be attributed to a very tight fit of the device that could be achieved in the KEMAR, whereas the additional probe tube unavoidably generated an additional small leak in the human subjects. Systematic differences to human data occurring beyond 10 kHz, which were most prominent in data including the response of external sound sources at the occluded eardrum, were probably caused by insufficient SNR in the human data. The KEMAR with anthropometric pinnae thus seems like a suitable tool to assess the acoustic properties of in-ear hearing devices in a median human ear.

\section*{Data Availability}
The database is available under the Creative Commons CC-BY-SA 4.0 licence at \url{https://doi.org/10.5281/zenodo.3733191}. The data is supplied both in a custom MAT format for use in matlab/octave as well as the dedicated SOFA format.

\section*{Acknowledgement}
This reserach was funded by the Deutsche Forschungsgemeinschaft (DFG) – Projektnummer 352015383 – SFB 1330 HAPPAA A4.
We acknowledge Meike Renken for placing the probe tubes and executing most experiments with the human subjects, Bernhard Eurich and Christoph Kirsch for assistance, and our motivated subjects for enduring these measurements keeping their heads still.

\small

\bibliographystyle{unsrt}

\begin{thebibliography}{33}
	\providecommand{\natexlab}[1]{#1}
	\providecommand{\url}[1]{\texttt{#1}}
	\expandafter\ifx\csname urlstyle\endcsname\relax
	\providecommand{\doi}[1]{doi: #1}\else
	\providecommand{\doi}{doi: \begingroup \urlstyle{rm}\Url}\fi
	
	\bibitem{grimm_toolbox_2019}
	Giso Grimm, Joanna Luberadzka, and Volker Hohmann.
	\newblock A {Toolbox} for {Rendering} {Virtual} {Acoustic} {Environments} in the {Context} of {Audiology}.
	\newblock \emph{Acta Acustica united with Acustica} 105 (2019) 566--578.
	\newblock \doi{10.3813/AAA.919337}.
	
	\bibitem{pausch_extended_2018}
	Florian Pausch, Lukas Asp{\"o}ck, Michael Vorl{\"a}nder, and Janina Fels.
	\newblock An {Extended} {Binaural} {Real}-{Time} {Auralization} {System} {With} an {Interface} to {Research} {Hearing} {Aids} for {Experiments} on {Subjects} {With} {Hearing} {Loss}.
	\newblock \emph{Trends in Hearing} 22 (2018) 2331216518800871.
	\newblock \doi{10.1177/2331216518800871}.
	
	\bibitem{denk_adapting_2018}
	Florian Denk, Stephan M.~A. Ernst, Stephan~D. Ewert, and Birger Kollmeier.
	\newblock Adapting {Hearing} {Devices} to the {Individual} {Ear} {Acoustics}: {Database} and {Target} {Response} {Correction} {Functions} for {Various} {Device} {Styles}.
	\newblock \emph{Trends in Hearing} 22 (2018) 2331216518779313.
	\newblock \doi{10.1177/2331216518779313}.
	
	\bibitem{oreinos_measurement_2013}
	Chris Oreinos and J{\"o}rg~M. Buchholz.
	\newblock Measurement of a {Full} {3D} {Set} of {HRTFs} for {In}-{Ear} and {Hearing} {Aid} {Microphones} on a {Head} and {Torso} {Simulator} ({HATS}).
	\newblock \emph{Acta Acustica united with Acustica} 99 (2013) 836--844.
	\newblock \doi{10.3813/AAA.918662}.
	
	\bibitem{kayser_database_2009}
	Hendrik Kayser, Stefan Ewert, J{\"o}rn Anem{\"u}ller, Thomas Rohdenburg, Volker Hohmann, and Birger Kollmeier.
	\newblock Database of {Multichannel} {In}-{Ear} and {Behind}-the-{Ear} {Head}-{Related} and {Binaural} {Room} {Impulse} {Responses}.
	\newblock \emph{EURASIP Journal on Advances in Signal Processing}
	1 (2009) 298605.
	\newblock \doi{10.1155/2009/298605}.
	
	\bibitem{denk_spectral_2018}
	Florian Denk, Stephan~D. Ewert, and Birger Kollmeier.
	\newblock Spectral directional cues captured by hearing device microphones in individual human ears.
	\newblock \emph{The Journal of the Acoustical Society of America} 14 (2018) 2072--2087.
	\newblock \doi{10.1121/1.5056173}.
	
	\bibitem{durin_acoustic_2014}
	Virginie Durin, Simon Carlile, Pierre Guillon, Virginia Best, and Sridhar Kalluri.
	\newblock Acoustic analysis of the directional information captured by five different hearing aid styles.
	\newblock \emph{The Journal of the Acoustical Society of America} 136 (2014) 818--828.
	\newblock \doi{10.1121/1.4883372}.
	
	\bibitem{hoffmann_quantitative_2013}
	Pablo Hoffmann, Flemming Christensen, and Dorte Hammersh{\o}i.
	\newblock Quantitative assessment of spatial sound distortion by the semi-ideal recording point of a hear-through device.
	\newblock In \emph{Proceedings of {Meetings} on {Acoustics}} 19 (2013) 050018.
	\newblock \doi{10.1121/1.4799631}.
	
	\bibitem{moore_personalized_2019}
	Alastair~H. Moore, Jan~Mark de~Haan, Michael~Syskind Pedersen, Patrick~A.
	Naylor, Mike Brookes, and Jesper Jensen.
	\newblock Personalized signal-independent beamforming for binaural hearing aids.
	\newblock \emph{The Journal of the Acoustical Society of America} 145 (2019) 2971--2981.
	\newblock \doi{10.1121/1.5102173}.
	
	\bibitem{dillon_hearing_2012}
	Harvey Dillon.
	\newblock \emph{Hearing {Aids}}.
	\newblock Boomerang Press, Turramurra, 2nd edition, 2012.
	
	\bibitem{denk_equalization_2018}
	Florian Denk, Henning Schepker, Simon Doclo, and Birger Kollmeier.
	\newblock Equalization filter design for achieving acoustic transparency in a semi-open fit hearing device.
	\newblock In \emph{Proc. 13. {ITG} {Conference} on {Speech} {Communication}}
	(2018) 226--230, Oldenburg, Germany.
	
	\bibitem{valimaki_assisted_2015}
	Vesa V{\"a}lim{\"a}ki, A.~Franck, Jussi R{\"a}m{\"o}, H.~Gamper, and
	L.~Savioja.
	\newblock Assisted {Listening} {Using} a {Headset}: {Enhancing} audio perception in real, augmented, and virtual environments.
	\newblock \emph{IEEE Signal Processing Magazine} 32 (2015)
	92--99.
	\newblock \doi{10.1109/MSP.2014.2369191}.
	
	\bibitem{spriet_feedback_2008}
	Ann Spriet, Simon Doclo, Marc Moonen, and Jan Wouters.
	\newblock Feedback {Control} in {Hearing} {Aids}.
	\newblock In \emph{Springer {Handbook} of {Speech} {Processing}}, Springer {Handbooks}, Jacob Benesty, M.~Mohan Sondhi, and Yiteng~Arden Huang, Editors,. Springer Berlin Heidelberg, 2008.
	\newblock \doi{10.1007/978-3-540-49127-9_48}.
	
	\bibitem{blau_acoustics_2008}
	Matthias Blau, Tobias Sankowsky, Alfred Stirnemann, Hannes Oberdanner, and Nicola Schmitt.
	\newblock Acoustics of open fittings.
	\newblock \emph{The Journal of the Acoustical Society of America}, 123 (2008) 3011.
	\newblock \doi{10.1121/1.2932603}.
	
	\bibitem{denk_one-size-ts-all_2019}
	Florian Denk, Miriam Lettau, Henning Schepker, Simon Doclo, Reinhild Roden,  Matthias Blau, J{\"o}rg-Hendrik Bach, Jan Wellmann, and Birger Kollmeier.
	\newblock A one-size-fits-all earpiece with multiple microphones and drivers for hearing device research.
	\newblock In \emph{Proc. 2nd {AES} {Conference} on {Headphone} {Technology}} Paper 13 (2019) 1--9, San Francisco, USA.
	
	\bibitem{wille_iec_2016}
	Morten Wille and Per Rasmussen.
	\newblock {IEC} 60318-4 {Ear} {Simulator} for {Low} {Noise} {Measurements} \& {Anthropometric} {Rubber} {Pinna}.
	\newblock In \emph{Proc. {AES} {Conference} on {Headphone} {Technology}} (2016) 96--102, Aalborg, Denmark.
	
	\bibitem{schepker_null-steering_2019}
	H.~Schepker, S.~E. Nordholm, L.~T.~T. Tran, and S.~Doclo.
	\newblock Null-{Steering} {Beamformer}-{Based} {Feedback} {Cancellation} for {Multi}-{Microphone} {Hearing} {Aids} {With} {Incoming} {Signal}
	{Preservation}.
	\newblock \emph{IEEE/ACM Transactions on Audio, Speech, and Language
		Processing} 27 (2019) 679--691.
	\newblock \doi{10.1109/TASLP.2019.2892234}.
	
	\bibitem{liebich_active_2016-1}
	Stefan Liebich, Carlotta Anemuller, Peter Vary, Peter Jax, Daniel Ruschen, and Steffen Leonhardt.
	\newblock Active noise cancellation in headphones by digital robust feedback control.
	\newblock In \emph{Proc. 24th {European} {Signal} {Processing} {Conference} ({EUSIPCO})} (2016) 1843--1847, Budapest, Hungary.
	\newblock \doi{10.1109/EUSIPCO.2016.7760567}.
	
	\bibitem{vogl_transfer_2018}
	Steffen Vogl and Matthias Blau.
	\newblock Transfer functions in the ear canal depending on different sources ({Transferfunktionen} im {Geh{\"o}rgang} in {Abh{\"a}ngigkeit} verschiedener {Quellen}).
	\newblock In \emph{Fortschirtte der {Akustik} - {DAGA}} (2018) 661--614, Munich, Germany.
	
	\bibitem{rumsey_headphone_2019}
	Francis Rumsey.
	\newblock Headphone {Technology}: {Hear}-through, bone conduction, noise
	canceling.
	\newblock \emph{Journal of the Audio Engineering Society} 67 (2019) 914--919.
	
	\bibitem{pavlovic_open_2019}
	Caslav Pavlovic, Hendrik Kayser, Paul Maanen, Tobias Herzke, Volker Hohmann,
	S.~R. Prakash, and Reza Kasayan.
	\newblock Open {Portable} {Platform} for {Hearing} {Aid} {Research}.
	\newblock \emph{Presented at the {American} {Auditory} {Society} {Meeting} ({AAS})} (2019) Scottsdale, USA.
	
	\bibitem{herzke_open_2017}
	Tobias Herzke, Hendrik Kayser, Frasher Loshaj, Giso Grimm, and Volker Hohmann.
	\newblock Open signal processing software platform for hearing aid research ({openMHA}).
	\newblock In \emph{Proceedings of the {Linux} {Audio} {Conference}}, (2017) 35--42, Saint-Etienne, France.
	
	\bibitem{hoertech_technical_2019}
	Hoertech, InEar, and University of~Oldenburg.
	\newblock Technical documentation: {The} {Hearpiece}, a one-size-fits all
	in-ear research hearing device, 2019.
	\newblock URL
	\url{https://www.hoertech.de/images/hoertech/pdf/fe-produkte/TransparentEarPiece_Techdoc_v01.pdf}.
	
	\bibitem{denk_controlling_2017}
	Florian Denk, Jan Heeren, Stephan~D. Ewert, Birger Kollmeier, and Stephan M.~A.  Ernst.
	\newblock Controlling the {Head} {Position} during individual {HRTF} {Measurements} and its {Effect} on {Accuracy}.
	\newblock In \emph{Fortschritte der {Akustik} - {DAGA}} (2017) 1085--1088, Kiel, Germany.
	
	\bibitem{majdak_multiple_2007}
	Piotr Majdak, Peter Balazs, and Bernhard Laback.
	\newblock Multiple exponential sweep method for fast measurement of
	head-related transfer functions.
	\newblock \emph{Journal of the Audio Engineering Society} 55 (2007) 623--637.
	
	\bibitem{denk_removing_2018}
	Florian Denk, Birger Kollmeier, and Stephan Ewert.
	\newblock Removing reflections in semianechoic impulse responses by
	frequency-dependent truncation.
	\newblock \emph{Journal of the Audio Engineering Society} 66 (2018) 146--153.
	\newblock \doi{10.17743/jaes.2018.0002}.
	
	\bibitem{muller_transfer-function_2001}
	Swen M{\"u}ller and Paulo Massarani.
	\newblock Transfer-function measurement with sweeps.
	\newblock \emph{Journal of the Audio Engineering Society} 49 (2001) 443--471.
	
	\bibitem{shaw_sound_1968}
	E.~A.~G. Shaw and R.~Teranishi.
	\newblock Sound {Pressure} {Generated} in an {External} {Ear} {Replica} and
	{Real} {Human} {Ears} by a {Nearby} {Point} {Source}.
	\newblock \emph{The Journal of the Acoustical Society of America} 44 (1968) 240--249.
	\newblock \doi{10.1121/1.1911059}.
	
	\bibitem{burkhard_anthropometric_1975}
	M.~D. Burkhard and R.~M. Sachs.
	\newblock Anthropometric manikin for acoustic research.
	\newblock \emph{The Journal of the Acoustical Society of America} 58 (1975) 214--222.
	\newblock \doi{10.1121/1.380648}.
	
	\bibitem{vogl_individualized_2019}
	Steffen Vogl and Matthias Blau.
	\newblock Individualized prediction of the sound pressure at the eardrum for an earpiece with integrated receivers and microphones.
	\newblock \emph{The Journal of the Acoustical Society of America} 145 (2019) 917--930.
	\newblock \doi{10.1121/1.5089219}.
	
	\bibitem{hoffmann_insert_2013}
	Pablo Hoffmann, Flemming Christensen, and Dorte Hammersh{\o}i.
	\newblock Insert {Earphone} {Calibration} for {Hear}-{Through} {Options}.
	\newblock In \emph{Proc. {Audio} {Engineering} {Society} {Conference} 51:
		{Loudspeakers} and {Headphones}} (2013) 1--8, Helsinki, Finland.
	
	\bibitem{denk_individualised_2018}
	Florian Denk, Marko Hiipakka, Birger Kollmeier, and Stephan M.~A. Ernst.
	\newblock An individualised acoustically transparent earpiece for hearing
	devices.
	\newblock \emph{International Journal of Audiology} 57 (2018) 62--70.
	\newblock \doi{10.1080/14992027.2017.1294768}.
	
	\bibitem{roden_using_2020}
	Reinhild Roden, Nick Wulbusch, Alexey Chernov, Florian Denk, and Matthias Blau.
	\newblock Using an electro-acoustic model of a vented earpiece to predict the
	ear canal input impedance.
	\newblock In \emph{Submitted to {Forum} {Acusticum}} (2020)  1--6, Lyon, France.
	
\end{thebibliography}

\end{document}